# Time-dependent neo-deterministic seismic hazard scenarios: Preliminary report on the M6.2 Central Italy earthquake, 24[th] August 2016

Antonella Peresan [1,2,4], Vladimir Kossobokov [3,4], Leontina Romashkova [3], Andrea Magrin [2], Alexander Soloviev [3], Giuliano F Panza [4,5]

1 - CRS-OGS, National Institute of Oceanography and Experimental Geophysics. Udine. Italy
2 - Department of Mathematics and Geosciences, University of Trieste – Italy
3 - IEPT, Russian Academy of Sciences, Moscow, Russian Federation
4 - International Seismic Safety Organization (ISSO)
5 - Institute of Geophysics, China Earthquake Administration, Beijing

## 1. Introduction

A reliable and comprehensive characterization of expected seismic ground shaking, eventually including the related time information, is essential in order to develop effective mitigation strategies and increase earthquake preparedness. Forecasting earthquakes and related ground shaking, however is not an easy task and it implies a careful application of statistics to data sets of limited size and different accuracy. Nowadays it is well recognized by the engineering community that standard hazard indicator estimates (e.g. seismic PGA) alone are not sufficient for the adequate design, mainly for special buildings and infrastructures. Moreover, any effective tool for SHA must demonstrate its capability in anticipating the ground shaking related with large earthquake occurrences, a result that can be attained only through rigorous verification and validation process.

A scenario-based Neo-Deterministic approach to Seismic Hazard Assessment (NDSHA) is available nowadays, which considers a wide range of possible seismic sources (including the largest deterministically or historically defined credible earthquake, MCE) as the starting point for deriving scenarios by means of full waveforms modeling, either at national and local scale. The method does not make use of attenuation relations and naturally supplies realistic time series of ground shaking, including reliable estimates of ground displacement readily applicable to seismic isolation techniques. The NDSHA procedure permits to incorporate, as they become available, new geophysical and geological data, leading to the natural definition of a set of scenarios of expected ground shaking at the bedrock. At the local scale, further investigations can be performed taking into account the local soil conditions, in order to compute the seismic input (realistic synthetic seismograms) for engineering analysis of relevant structures, such as historical and strategic buildings. The standard NDSHA has been already applied in several regions worldwide, including a number of local scale studies accounting for two-dimensional and three-dimensional lateral heterogeneities in anelastic media.

Based on the neo-deterministic approach, an operational integrated procedure for seismic hazard assessment has been developed that allows for the definition of time dependent scenarios of ground shaking, through the routine updating of earthquake predictions, performed by means of the algorithms CN and M8S (Peresan et al., 2005). The integrated NDSHA procedure for seismic input definition, which is currently applied to the Italian territory, combines different pattern recognition techniques, designed for the space-time identification of strong earthquakes, with algorithms for the realistic modeling of ground motion. Accordingly, a set of deterministic scenarios of ground motion at bedrock, which refers to the time interval when a strong event is likely to occur within the alerted area, can be defined by means of full waveform modeling, both at regional and local scale. CN and M8S predictions, as well as the related time-dependent ground motion scenarios associated with the alarmed areas, are routinely updated every two months since 2006 (Panza et al., 2012; Peresan et al., 2011).

## 2. Intermediate-term middle-range earthquake predictions by CN and M8S algorithms

CN and M8S predictions, as well as the related time-dependent ground motion scenarios associated with the alarmed areas, are routinely updated every two months since 2006. The rules for the real-time application of CN and M8S algorithms to the Italian territory are described in detail in Peresan et al. (2005), whereas the procedure for the definition of the related ground shaking scenarios is illustrated in Peresan et al. (2011).

The intermediate-term middle-range earthquake prediction experiment, aimed at a real-time testing of M8S and CN predictions for earthquakes with magnitude larger than a given threshold (namely 5.4 and 5.6 for CN algorithm, and 5.5 for M8S algorithm) in the Italian region and its surroundings, is ongoing since 2003. Predictions are regularly updated every two months and a complete archive of predictions is made available on-line (http://www.geoscienze.units.it/esperimento-di-previsione-dei-terremoti-mt.html), thus allowing for a rigorous validation of the applied algorithms. The results obtained during more than nine years of real-time monitoring already permitted a preliminary assessment of the significance of the issued predictions (Peresan et al., 2011). So far, 14 out of the 16 strong earthquakes, occurred within the monitored territory since 1954, have been correctly preceded by an alarm (TIP, Time of Increased Probability) declared by CN algorithm, with about 30% of the overall space–time volume occupied by alarm[1]; the confidence level for such predictions is above 99%. Similarly, the algorithm M8S correctly identified 14 of the 23 earthquakes with magnitude M5.5+ (i.e., between 5.5 and 6.0), occurred since 1972 within the monitored territory, with a space–time volume of alarm of about 31%; the confidence level of M5.5+ predictions has been estimated to be above 98% (no estimation is yet possible for higher magnitude levels).

A strong earthquake (M=6.2) hit the Rieti province, Central Italy, on 24th August 2014. The epicenter was located inside the Central Region (Fig. 1), alerted by CN algorithm for an earthquake with magnitude M≥5.6, starting on 1 November 2012, whereas it occurred outside the areas alerted by M8S algorithm for the corresponding magnitude interval (Fig. 2). Therefore the earthquake scores as a successful real-time prediction, for CN algorithm only. The epicenter of the event falls outside the area alerted by M8S for an earthquake with magnitude 6.0≤M<6.5, as on August 2016 (Fig. 2b); however, the epicentral area was in state of alarm less than one year before, i.e. up to December 2015 (Fig. 2a). During 2016 the alarm area shrunk down to the south, thus resulting in a failure to predict.

The uncertainties associated with intermediate-term middle-range earthquake predictions are intrinsically quite large. CN and M8S algorithms, however, already proved effective in predicting strong earthquakes, by rigorous prospective testing over the Italian territory. Specifically, 6 out of 8 target earthquakes have been predicted in real-time by CN, and 5 out of 9 by M8S algorithm. For both the algorithms the confidence level achieved in real-time testing is above 97%. From the diagram of alarms (TIPs) reported in figure 1, it is possible to observe that 5 out of 8 alarms where followed by an earthquake, with a false alarm rate around 35% for CN Central region.

---

[1] The prediction experiment by CN algorithm has been recently expanded to the Adria Region, where predictions are routinely updated since 2005 (Peresan, 2016 - Chapter submitted to AGU Book on "Pre-Earthquake Processes"). When including the Adria Region the score increases to 21 out of 25 strong earthquakes correctly diagnosed by CN algorithm, with about 31% of TIPs, and a confidence level well above 99%.

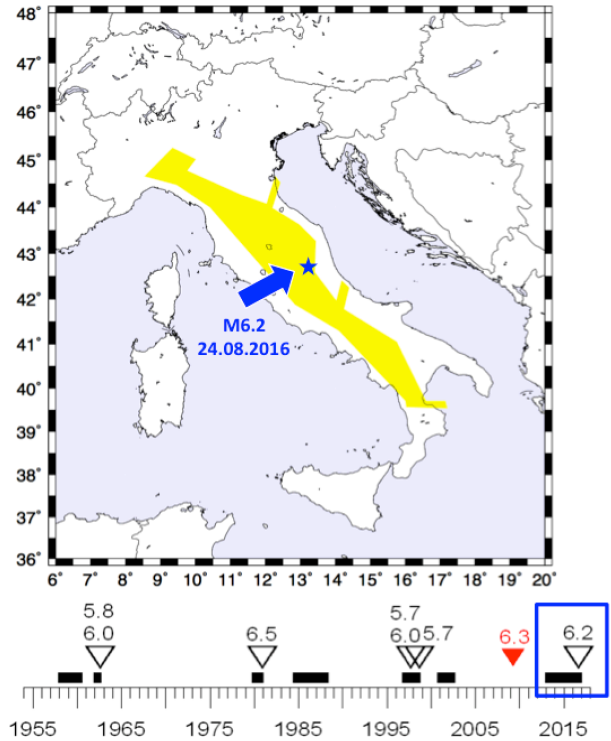

Fig. 1 – Central Region (in yellow) alerted by CN algorithm for the time interval 1.11.2012-1.5.2017. In the diagram of TIPs, the black boxes represent periods of alarm, while each triangle surmounted by a number indicates the occurrence of a strong event (M≥$M_0$=5.6), together with its magnitude. Full red triangles indicate failures to predict. The blue star in the map and the blue arrow indicate the August 24$^{th}$ 2016 Central Italy earthquake. The complete predictions archive is available at: www.geoscienze.units.it/esperimento-di-previsione-dei-terremoti-mt.html.

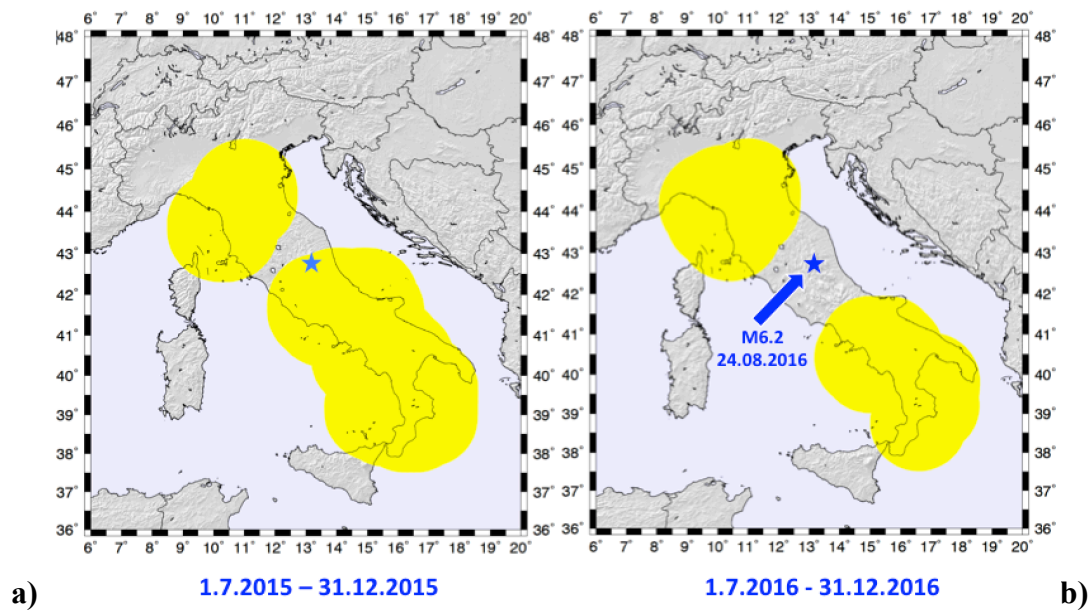

a)   1.7.2015 – 31.12.2015                    1.7.2016 - 31.12.2016   b)

Fig. 2 – Areas alerted by M8S algorithm (in yellow) for a possible earthquake with magnitude 6.0≤M<6.5 determined for two time windows shifted by one year, namely: a) 1.7.2015 – 31.12.2015 and a) 1.7.2016 – 31.12.2016. The blue stars mark the epicenter of the August 24$^{th}$ 2016 earthquake.

## 3. Neo-deterministic time-dependent seismic hazard scenarios for the Italian territory

Based on NDSHA approach, an operational integrated procedure for seismic hazard assessment has been developed (Peresan et al., 2011) that allows for the definition of time dependent scenarios of ground shaking, through the routine updating of earthquake predictions, performed by means of the algorithms CN and M8S (Peresan et al., 2005). The integrated NDSHA procedure for seismic input definition, which is currently applied to the Italian territory, combines different pattern recognition techniques, designed for the space-time identification of strong earthquakes, with algorithms for the realistic modeling of ground motion. Accordingly, a set of deterministic scenarios of ground motion at bedrock, which refers to the time interval when a strong event is likely to occur within the alerted areas can be defined by means of full waveform modeling, both at regional and local scale.

In Italy and surrounding regions the areas prone to strong earthquakes have been systematically identified based on the morphostructural zonation and pattern recognition analysis, considering two magnitude thresholds (M$\geq$6.0 and M$\geq$6.5) as described by Gorshkov et al. (2002; 2004). The identified seismogenic nodes are used, along with the seismogenic zones ZS9 (Meletti and Valensise, 2004), to characterize the earthquake sources used in the seismic ground motion modeling, as described by (Peresan et al., 2011). The earthquake epicentres reported in the catalogue are grouped into 0.2°x0.2° cells, assigning to each cell the maximum magnitude recorded within it. A smoothing procedure is then applied, to account for spatial uncertainty and for source dimensions. Only the sources located within the alarmed areas are considered to define the time-dependent scenarios (Panza et al., 2012; Peresan et al., 2011). From the set of complete synthetic seismograms, different maps that describe the maximum ground shaking at the bedrock can be produced, including peak ground displacement (PGD), velocity (PGV) or acceleration (PGA).

The August 24$^{th}$ 2016 Central Italy earthquake took place within one of the seismogenic nodes previously identified as prone to possible earthquakes with magnitude M$\geq$6.0 by Morphostructural Zonation and pattern recognition analysis (Gorshkov et al., 2002). This earthquake prone area experienced other destructive earthquakes in the past, including a M6.2 earthquake in 1639 and the large M6.9 Valnerina earthquake in 1703.

The seismic hazard maps at the bedrock defined by the Neo-Deterministic approach (NDSHA), correctly anticipated the recorded ground shaking. Specifically, in the NDSHA map published in 2001 (Panza et al., 2001) the predicted value is 0.15-0.3g, and the revision published in 2012 (Panza et al., 2012) gives 0.3-0.6g, which well contain the recorded values as high as 0.45g (RAN data: http://ran.protezionecivile.it/IT/dettaglio_evid.php?evid=340867). This is quite natural since the August 24 earthquake did not necessarily generate the largest possible shaking in the area, as evidenced by the M6.9 Valnerina earthquake that struck the area in 1703, and should be factored in reconstruction considering that source and local effects may lead to values >0.6g. Notably, although the earthquake occurred in a well known seismic region, the ground shaking for this event exceeded (about 50% higher) at several sites the values expected at the bedrock according to current Italian seismic regulation (i.e. PGA<0.275g), which is based on a classical PSHA map (Gruppo di Lavoro, 2004) and overlooks source and site specific conditions. In conclusion, even when grossly adjusted for soil type corrections, following the rules given in the current Italian building code (C.S.L.P. 2008 - Norme Tecniche per le Costruzioni, http://www.cslp.it/cslp/), PSHA still underestimates the maximum recorded ground shaking by about 30%. NDSHA estimates, instead, when accounting for soil conditions, following the same rules given by C.S.L.P. (2008), are well compatible with available recordings, and represent a serious warning to be considered in any future action.

The time-dependent ground shaking scenario associated to CN Central region (Fig. 3a) defined for the period 1 November 2012 – 1 September 2016, appears also well consistent with the ground shaking recorded for this earthquake. Since the time NDSHA time-dependent scenarios are regularly computed, namely starting on 2006, this is the third large earthquake that struck the Italian territory, along with L'Aquila earthquake. In all cases the method correctly predicted the observed ground motion, although L'Aquila earthquake scores as a failure in the earthquake prediction experiment, because the epicenter was located about ten km outside the alarmed territory.

The prospective application of the time-dependent NDSHA approach provides information that can be useful in assigning priorities for timely mitigation actions and, at the same time, allows for a rigorous prospective testing and validation of the proposed methodology. A broad spectrum of interrelated actions can be undertaken for mitigation of earthquake impact on cultural heritage, including temporary safety measures and long term (tens of years) planning of interventions and retrofitting (Vaccari et al., 2009).

The possible role of intermediate-term middle-range earthquake predictions in forecasting and planning safeguard interventions for cultural heritage, was discussed at the Conference "Resilienza delle città d'arte ai terremoti" ([Accademia Nazionale dei Lincei, Rome,](#) 3-4 November 2015). The above mentioned methodologies and results were recently presented both at the International Conference on "Data Intensive System Analysis for Geohazard Studies" (Sochi, 18–21 July 2016), as well as at the Moscow School for Young Scientists on "System Analysis and Seismic Hazard Assessment" (12-15 July 2016).

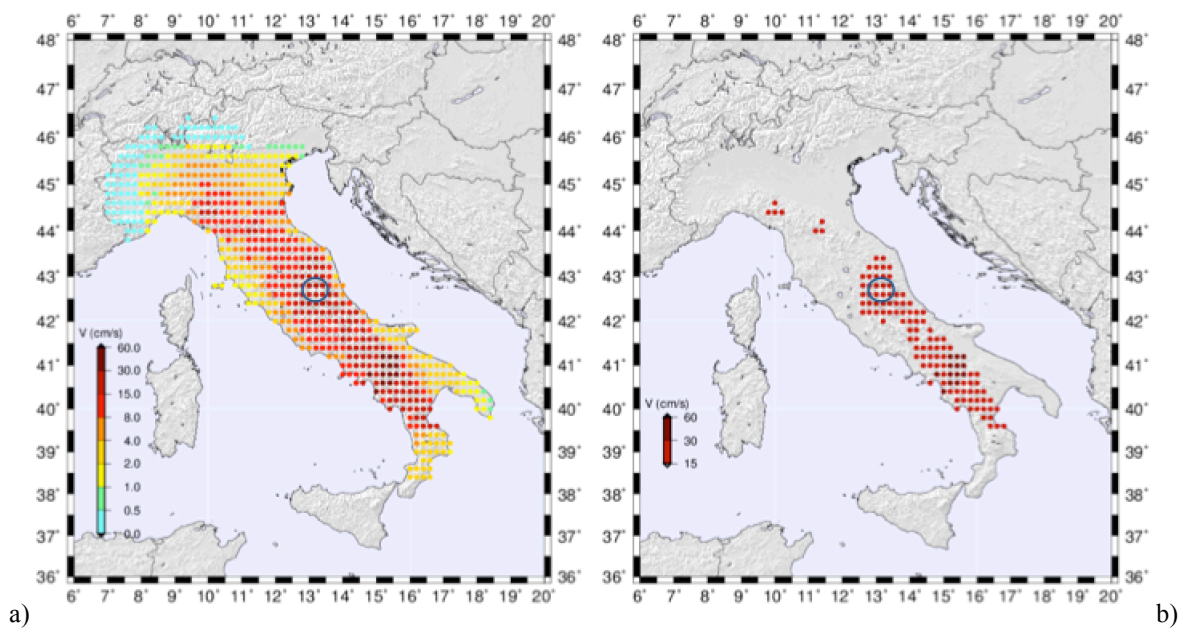

Fig. 3 –Time-dependent scenarios of ground shaking associated to the alarm in CN Central Region (Fig. 1). On the left maps of peak ground velocity (PGV) are shown, computed using simultaneously all of the possible sources within the alarmed area and for frequencies up to 10 Hz. On the right, the same maps are provided, but for PGV>15 cm/s (modified after Peresan et al., 2015). The circle on maps a) and b) evidences the area within 30 km distance from the epicenter of the Central Italy earthquake (M6.2, 2016).